\documentclass[twocolumn,prl,showpacs,preprintnumbers]{revtex4}

\usepackage{graphics}
\usepackage{amssymb}
\usepackage{amsmath}
\usepackage[latin1]{inputenc}
\usepackage{natbib}
\usepackage{graphicx}
\usepackage{epsfig}
\begin{document}

\title{ Time Delays in the Synchronization \\of Chaotic Coupled Systems with Feedback}
\author{Jhon F. Martinez Avila and J. R. Rios Leite}
\affiliation{Departamento de F\'{\i}sica,~ Universidade Federal de Pernambuco, 50670-901, 
Recife, PE, Brazil
}

\date{\today}

\begin{abstract}

Isochrony and time leadership was studied in the synchronized excitable behavior 
of coupled chaotic lasers.
Each unit of the system had chaos due to feedback with a fixed delay time. 
The inter-units coupling signal had a second, independent, characteristic time.
Synchronized excitable spikes present isochronous, time leading or time lagging behavior whose stability is shown to depend on a simple relation between the feedback and the coupling times.
Experiments on the synchronized low frequency fluctuations of two optically coupled semiconductor lasers and numerical calculations with coupled laser equations verify the predicted 
stability conditions for synchronization.
Synchronism with intermittent time leadership exchange was also observed and characterized.

\end{abstract}


\maketitle
 
Many dynamical systems in nature have chaos due to feedback. A characteristic time, $\tau_F$,  
associated to the feedback is therefore imbedded in the system response. As two or more of such 
systems are coupled, another independent time, $\tau_C$, corresponding to the time of flight of 
the coupling signal, enters in the dynamical description of the global system. 
We present here how the relation between these two times determines 
the possible time delays in chaos synchronization. 
New features of time leadership competition in synchronized chaos between coupled pairs of systems 
with feedback are found when feedback and inter-coupling times have the same order of magnitude. 
Experimental and numerically, the stability of isochronous
 chaos synchronism for identical systems  occurs only for specific 
relations between these times.
We also show that time delayed and time advanced synchronism as well as
synchronism with
intermittent leadership exchange are also quantitatively 
determined by the ratio of these times. Our case is made with pairs of 
semiconductor diode lasers. However, the properties of synchronization in complex systems extends far beyond physical devices  \cite{kurthbook}, reaching the subject of neural sciences \cite{mirassoPNAS2008}.  

\begin{figure}[!phbt]
 \centering
\includegraphics[width=5.0cm,bb=0 0 658 570]{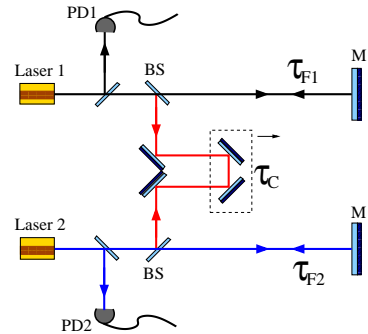}
 \caption{Experimental setup for the synchronization of Low Frequency Fluctuations in 
optically coupled laser }
 \label{fig:setup}
\end{figure}

The study of synchronism with chaotic lasers spreads for more than a decade  \cite{sugawara1994,roy1994,liu1994,meucci2006,kanter2007}.
Of relevant interest for applications are the results on the 
synchronization of semiconductor lasers \cite{takigushi1999,mirassonature2005} 
for encrypted 
communication. With optical feedback diode lasers can present 
chaos in the form of very fast output power fluctuations, at a time scale of picoseconds. 
Superimposed on these, irregular power drops occur in a much slower time scale (order of hundreds 
of nanoseconds and longer) corresponding to the so called low frequency fluctuations (LFF)  \cite{tartwijklenstra1995}. The LFF drops in single lasers are a current subject 
of studies and have been associated  
to spikes of excitable systems \cite{giudiciandronov1997,avilaPRL2004,avilaPRL2008}.
Coupled diode lasers show time advanced and time lagging synchronization via unidirectional coupling in master-slave configuration \cite{cristina2001,alan2001}  and in mutually coupled systems \cite{mirasso2001}. 
Symmetrically coupled pairs of lasers, without feedback, were shown to have unstable isochronous 
chaotic pulsation \cite{mirasso2001,mulet2004,dakin2006}. 
In the experiments and calculations with coupled lasers without feedback,
the time leading lasers always appears with its power drops displaced by one unit of
the coupling time, $\tau_C$.
 The use of intermediate relaying system
was demonstrated to give stable isochrony \cite{fischer2006,shaw2007}. Isochronous synchronization
has also been investigated \cite{klein2006,shaw2007,lev2006} for lasers with feedback with
the studies focused on the fast laser fluctuations. 
Differently, herein we study the synchronization of the low frequency fluctuations (LFF),
known to appear at the scale of hundreds of nanoseconds and slower.
Thus, in our case the 
dynamics in the coupled systems is 
much faster than the synchronized events which have the properties of excitable spikes. So, our results refer to synchronism of excitable dynamical systems.
The schema of the experiments is given in figure \ref{fig:setup}. 
Optical feedback was created in each laser by a retroreflecting external mirror. 
Their feedback return times, $\tau_{F1}$ and  $\tau_{F2}$, were set equal to within $1\%$ 
precision (both named from hereon  $\tau_F$). 
The time of travel of the coupling signal between the lasers, $\tau_C$, was independently
controlled  with respect to $\tau_F$. 
Small changes in either of the times, on the scale of fraction of nanoseconds
do not alter the properties of the synchronism. Thus the results, 
like the LFF phenomenon in single laser with feedback, 
are robust with respect to optical phase changes.
A consequence of such behavior is that our observations are consistent with the
 synchronization of excitable systems \cite{ciszakMexcitable2004}. These authors show how 
unidirectionally coupled excitable systems can presents one of the systems always with a lower threshold for excitability. Their coupled systems synchronize to an external common signal. In our case the coupled systems have fast fluctuations and so, no external source of excitation is necessary.

\begin{figure}[!phbt]
 \centering
 \includegraphics[width=8.0cm]{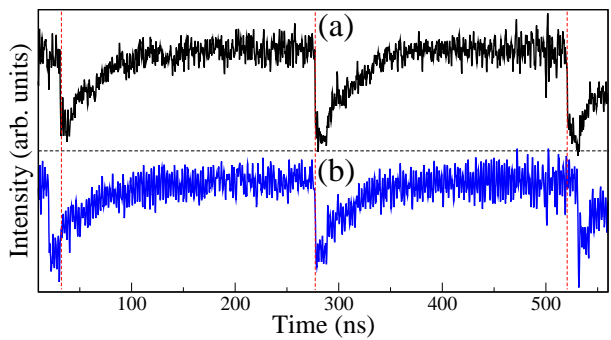}
 \caption{Segment of experimental time series of the power of (a) laser 1 and (b) 
laser 2, showing the 
low frequency synchronism and including the interchange of time delay between the lasers.
$\tau_{F}$ = 10 ns and $\tau_{C}$ = 10 ns.
}
 \label{fig:expseriesync10ns}
\end{figure}

The experiments were done with two SDL 5401 {\it GaAlAs} semiconductor lasers, named here as
Laser 1 and Laser 2, both with solitary threshold current of $21$ mA and emitting at 805 nm.
They were thermally stabilized
to $0.01$ K and could be temperature tuned to have their optical frequencies 
within $2$ Ghz separation.
The feedback time of the lasers was  $\tau_F$ = $10 \, ns$ and the strength
were measured by the threshold reduction parameter,
$\xi$  which is the percentual variation of the threshold pump current as we consider the laser with and without feedback \cite{tartwijklenstra1995}.
Symmetrical optical coupling between the lasers was produced with $50\%$ beam splitters 
as shown in figure \ref{fig:setup}. The time of flight of the light between the lasers
was varied between $5$ and $20$ ns.
 The threshold reductions due to cross input power were used to quantify the
coupling strengths whose contribution to our studies will be detailed elsewhere.
Manipulating the laser currents and temperatures, LFF synchronism was obtained 
where to each power drop in laser 1 corresponds a 
drop in laser 2 and vice versa.
Typical experimental segments of the power output of the two lasers, with three events of
the synchronized low frequency fluctuations (LFF) drops, are
shown in figure~\ref{fig:expseriesync10ns}.
Each laser intensity was detected by a 1.5 GHz bandwidth photodiode.
Simultaneous  data series 
were acquired with a two channel digital oscilloscope having a bandwidth of 1 GHz 
and a maximum sampling rate of 5 GSamples/second.
  One of the lasers could  always be the time leader by setting 
its pump current or its optical frequency \cite{mirasso2001} higher.
 However, as careful tuning of
the laser currents and optical frequencies is made, within the same long LFF
synchronized time series the LFF drops of laser 1
can appear isochronous, leading or lagging these of laser 2.
Such were the conditions of figure~\ref{fig:expseriesync10ns} where the 
time leadership is interchanged between drops one and three, while the 
second LFF drops are isochronous.
Using  long pairs of data
series, coarse grained to a $1\,ns$ time resolution \cite{avilaPRL2008}, we measured
 $\Delta T_{i}$= $T_{1i}-T_{2i}$, the delay between 
the  LFF drop $i$ of laser 1 and laser 2.

The main result of this work is to show that, within any LFF synchronized
evolution, for any pair of excited drops the allowed delays are given by
\begin{equation}
 \Delta T_{i}=  m_i\cdot\tau_{C}-n_i\cdot\tau_{F}
\label{eq:delay}
\end{equation}
where $m_i\neq 0$ and $n_i$ are small integers. 
Here we only give indications of $m_i=\pm1$ but preliminary results with 
very long numerical series show rare events with $m_i>1$. 
Equation \ref{eq:delay} covers various previously studied cases in 
the literature and confirms 
novel observations.
For instance, with the lasers having no feedback, $\tau_{F}=0$,  it
verified that isochrony is not allowed \cite{mirasso2001}.
Other cases with feedback but forbidding isochrony are given below.
The equation determines $\Delta T_{i}$ when one laser is
always leading or lagging which happens for a single pair $(m_i,n_i)$ 
through the whole time series.
Furthermore, it is also valid for cases with varying values of $(m_i,n_i)$ within the 
same series, obtained from symmetrical  and nearly symmetrical systems. The LFF 
synchronism then appears with intermittent 
switching of the time delay that can interchange the leading subsystem.
The evidence of equation \ref{eq:delay} as a property of the LFF synchronism was
obtained inspecting many different
feedback and coupling delay times in experimental and numerical series.
The origin of equation \ref{eq:delay} is present in the intensity cross-correlation functions of the fast fluctuations of coupled lasers. These correlations, obtained with sub-nanosecond resolution, show recurrent narrow  spikes (width of hundred of picoseconds) 
at positions and intervals  given by \ref{eq:delay}. 

Let us proceed presenting our experiments along with the numerical-theoretical 
results extracted from
a system of differential equations that 
describe two mono-mode lasers with optical feedback and optical coupling.
The model correspond to a set of modified  \citet{langkobayashi} equations for the lasers
with feedback, including symmetrical optical coupling and assumed to 
have the same optical frequency:
\begin{eqnarray}
\dfrac{dE_{i}(t)}{dt} & = & \dfrac{\left( 1+i \alpha \right)}{2} \left[G(N_{i})-
\frac{1}{\tau_{p}}\right]E_{i}(t) +  \kappa E_{i}(t- \tau_{F})                                                                    \nonumber \\
                 &    &   + 
\gamma E_{j\neq i}(t-\tau_{C})\\
\dfrac{dN_{i}(t)}{dt} & = & J_i - \dfrac{N_{i}(t)}{\tau_{s}}- G(N_{i}) \left|E_{i}(t)\right|^{2},
\label{eq:LKEQS}
\end{eqnarray}
where $i,j=1,2$  and each laser gain is given by
\begin{equation}
G(N_{i}) = \frac{G_{0}(N_{i}-N_{0})}{1+ \epsilon \left|E_{i}(t)\right|^{2} }
\end{equation}
The definition of the various parameters and their typical values are well 
discussed in the 
literature \cite{tartwijklenstra1995,avilaPRL2004}. For each laser, $E_{j}$ is the radiation 
field, $\alpha$ is the factor describing amplitude to phase conversion, 
 G is the amplifying gain, $N_{j}(t)$ the carriers inversion population,  
$\tau_p$ the 
photon lifetime of the internal laser cavity, $\tau_s$ the carriers 
lifetime, $J_{j}$ the threshold normalized pump 
currents and $N_0$ 
the inversion population for medium transparency.
 Each feedback field
has an amplitude coefficient $\kappa$ and feedback time $\tau_F$.
The  optical couplings 
are linearly additive $E$ field with coefficient $\gamma$
and time $\tau_C$ for the field of one laser to reach the other one.
Physical causality demands that both $\tau_F$ and $\tau_C$ be positive.

\begin{figure}[!hbtp]
\centering
\includegraphics[width=8.0cm]{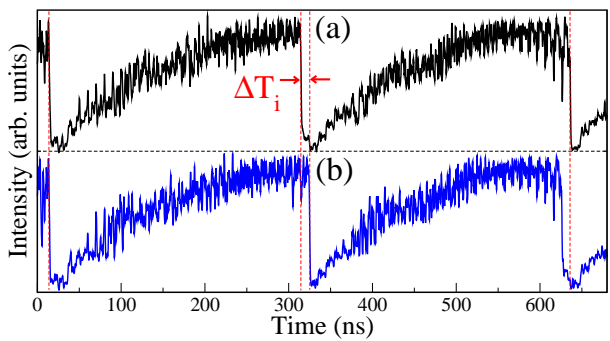}
\caption{Typical segment of the numerical time series of (a) laser 1 and (b) laser 2 LFF power 
drops under synchronized condition.
Notice that the first drops are isochronous while time leadership was interchanged 
between drops two and three. $\tau_{F}$ = 10 ns and $\tau_{C}$ = 10.05 ns. }
\label{fig:seriesLFF-LK10nsExcitable}
\end{figure}

 With a fourth order Runge-Kutta algorithm numerical data series were obtained 
for $E_{i}(t)$ and these gave the normalized intensity series $|E_{i}(t)|^2$.
 The time scales in the integrations and the equations parameters were fixed as:
$dt$=1 ps, $1/\tau_p$=282 ns$^{-1}$, 
$N_0 = 1.5\times10^{8}$, $\epsilon = 5\times10^{-7}$, 
$\kappa$=$\gamma$ = 22 ns$^{-1}$,
$1/\tau_s=1.66$ ns$^{-1}$. The  times, $\tau_{F}$ and $\tau_{C}$, are in the nanosecond range.
 Calculation with one laser having sufficiently higher pump current 
 gives synchronized 
LFF with the laser of higher pump current always leading in time, as observed in the experiments.
Equal pump currents ($J_i=1.013$), corresponding to symmetrical systems, produced numerical time series with leadership 
exchange, again in agreement with the experiment. Segments of these data series are shown in figure~\ref{fig:seriesLFF-LK10nsExcitable}, to be compared with the experimental segments in  figure~\ref{fig:expseriesync10ns}.
Let us emphasize that no stochastic terms are used in the equations.
The apparent random time distribution of the LFF
events in each laser as well as the time leadership switching in the synchronism 
is due to the high dimension deterministic chaos.

\begin{figure}[!phbt]
 \centering
 \includegraphics[width=8.0cm]{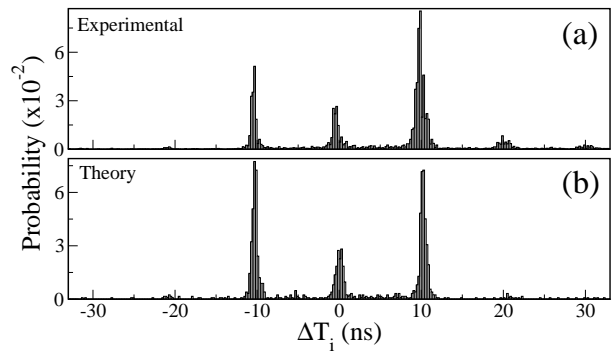}
 \caption{Histograms of the (a) experimental and (b) numerical time delays between LFF pulses 
of the two synchronized lasers. 
$\tau_{F}$ = 10 ns and $\tau_{C}$ = 10 ns.}
 \label{fig:hist-tcoup10ns}
\end{figure}

Time series with $\sim10^4$ pairs of drops were coarse grain filtered \cite{avilaPRL2008} 
with a $1$ ns time constant, and used 
 to extract histograms of time delays between the drops. 
 The switching of leadership from symmetrical systems
was also examined 
 and no correlation was found between consecutive values of  $\Delta T_i$, 
up to second order conditional probability.
This is indication of a Markov process, obtained despite the fact that the data came from  
deterministic numerical equations with recurrence times $\tau_F$ and $\tau_C$. 
Such result is consistent with high dimensional
 chaos with largely different time scales ($T_{i+1}-T_{i}\gg \tau_F, \tau_C$) that makes 
LFF appear as stochastic without memory \cite{avilaPRL2008}. The experimental data show 
the same lack of correlation for the delay times of LFF synchronism in symmetrical conditions.
The histograms associated to the data series of figures~\ref{fig:expseriesync10ns}
 and \ref{fig:seriesLFF-LK10nsExcitable}, 
are shown in figure~\ref{fig:hist-tcoup10ns}. 
As $\tau_{F}$ = $10$ ns and $\tau_{C}$ = $10$ ns, equation \ref{eq:delay} with $m_i=\pm1$ 
 and $n_i=\pm1$, predicts $\Delta T_i=0, \pm10\,ns$ and $\pm 20\,ns$. 
Indeed, the histograms show that 
isochrony occurs, along with time leadership exchange events.
The major probabilities are for events with 
 $\Delta T_{i} = 0$ and  $\pm$ 10 ns, with 
few events at $\pm20$ ns and some in $\pm30$ on the experimental data. 
It is important to mention that on the calculations small differences (less than 2\%) were introduced between the parameters of laser 1 and 2 in the equations. In particular $\tau_C\neq\tau_F$
($\tau_C=10.05$ ns) was necessary to give the non isochronous events. In fact  
such differences which are intrinsically present in the experiments 
change dramatically the amplitude of the peaks in the histograms 
but have minor effect on the
values of allowed delays. 
The robustness of the relation between delay times and the condition of LFF  
synchronization according to equation \ref{eq:delay} was also inspected for small, 
up to $10\%$, of relative variations of $\tau_{F}$  and $\tau_{C}$. 

\begin{figure}[!bthp]
\centering
\includegraphics[width=8.0cm]{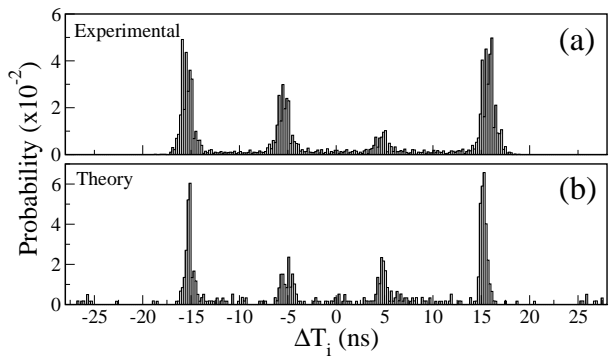}
\caption{Histograms from experimental (a) and numerical (b) series for the time delays 
between LFF drops of 
the two synchronized lasers: $\tau_{F}$ = 10 ns and $\tau_{C}$ = 15 ns.}
\label{fig:hist-tcoup15ns}
\end{figure}

Another typical numerical histogram where $m_i$ and $n_i$ change within 
the same synchronized 
evolution is shown in figure \ref{fig:hist-tcoup15ns}.
 This is also a case of symmetrical system where the laser parameters 
are equal but the values for time of coupling and feedback, 
 were chosen to give unstable isochrony: $\tau_{C} = 15$ ns and $\tau_{F}= 10$ ns.
Accordingly, equation  \ref{eq:delay} these values prevent $\Delta T_{LFF}=0$, as 
seen by direct substitution of small integers($m_i=\pm1$). Thus, synchronism of the LFF
holds, but there is always a finite time lead between the two lasers. 
The leadership can be exchange but
 events of simultaneous drops are excluded. 

\begin{figure}[!phbt]
\centering
 \includegraphics[width=8.0cm]{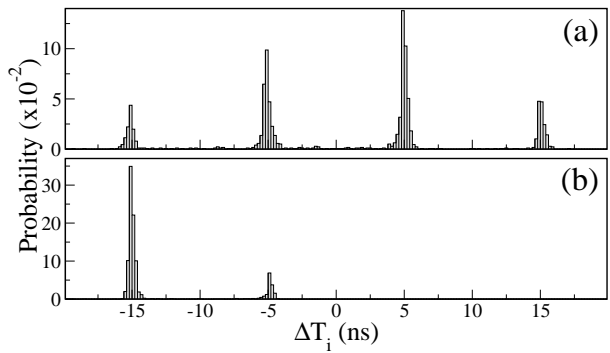}
 \caption{Histograms of the numerical delay times between synchronized LFF drops of the lasers
with  $\tau_{F} = 10$ ns and $\tau_{C}= 15$ ns and showing the effect of asymmetric conditions. In (a). $J_i = 1.013$ for the two lasers and in (b) $J_1 = 1.025$ and  $J_2 = 1.013$, making Laser 1 the time leader.}
 \label{fig:hist-tcoup15ns-LK.assim}
\end{figure}

When $\tau_{C} \gg \tau_{F}$ we verified numerically that equation \ref{eq:delay} still holds.
The value of $m_i$ is always $\pm1$ while $n_i$ assumes a large range of values. 
Isochrony is absent and the dominant events occur with $n_i=0$, corresponding 
to delays of $\pm \tau_C$. 
 Nonsymmetrical systems, as mentioned above, also follow equation \ref{eq:delay}. 
A calculation with the two lasers having the same current $J_i = 1.013$ was given figure \ref{fig:hist-tcoup15ns-LK.assim} (a) while figure \ref{fig:hist-tcoup15ns-LK.assim} 
(b) shows what happens when Laser 1 has its current augmented to $J_1 = 1.025$. 
 The laser with higher current is always the time 
leader, even though sometimes its leading time jumps down from $\tau_C$ 
to $\tau_C -\tau_F$.
These events correspond to be $m_i=1$ and $n_i=0$ changing to $m_i=1$ and $n_i=+1$ along the dynamical evolution with synchronized LFF excitations.

To summarize, from experiments with coupled lasers with feedback, corroborated by 
the numerical solutions of the corresponding rate equations, we discovered that 
the delay time in the synchronization of coupled excitable systems with feedback is 
controlled by a simple relation between feedback time and inter-coupling time. 
One of the  systems may have a fixed 
 time leadership or have its leading time switching values including isochrony.
 Nearly symmetrical systems can also have intermittent 
time leadership exchange always maintaining the excitable spikes synchronized in the large time scale. 
Specifically, the behavior of identical systems with feedback 
and coupled with time delay strongly depends on the relation between these times. 
A simple equation that specify the intercombinations of feedback and coupling time 
to give the allowed values of delay in the synchronized dynamics was introduced.
Coupled asymmetrical systems also follow the conditions for the allowed delay 
time between events. 
The various parameters of the systems strongly influence the 
probability of specific delays, distorting the histogram amplitudes, but have no effect on the values of the time delays.
Our results extend the previous  \cite{mirasso2001} determination of the 
symmetry breaking and instability of expected isochrony in the synchronism of 
identical coupled 
chaotic systems without feedback. It also adds to the recently studied 
properties of isochronal 
synchronism done by \citet{klein2006} and \citet{kanter2007}, who emphasize the 
potential application of these properties for encrypted communication. 
A formal mathematical treatment of the stable and 
unstable manifolds in chaotic synchronization with delays having integer 
combinations  between feedback and coupling 
time will be presented elsewhere. The phenomenon of selection condition in 
time delays is bound to appear generically in mutually coupled dynamical systems.

\acknowledgments {The authors acknowledge useful discussions with H. L. D. de S. Cavalcante.
Work partially supported by Brazilian
Agencies Conselho Nacional de Pesquisa e Desenvolvimento (CNPq) and
Funda\c{c}\~ao de Ci\^encia de Pernambuco (FACEPE) and by a Brazil-France 
Capes-Cofecub project 456/04.}
\bibliography{sync-refArXiv2009}
\end{document}